\documentclass[article, nojss]{jss}  
\graphicspath{{./}{figure/}{./}}

\usepackage{orcidlink,lmodern}
\usepackage[utf8]{inputenc}
\usepackage{mathtools, amssymb, amsthm, mathrsfs, amsmath}
\usepackage{placeins} 
\usepackage{framed}

\newcommand{\indep}{\perp\!\!\!\perp}

\author{Thomas Harder Scheike~\orcidlink{0000-0002-2148-4740}\\
  University at Copenhagen
  \AND
  Klaus Kähler Holst~\orcidlink{0000-0002-1364-6789}\\
  Novo Nordisk A/S}
\Plainauthor{Thomas Scheike, Klaus K\"ahler Holst}

\title{Restricted mean time lost for survival and competing risks data using
  \pkg{mets} in \proglang{R}}
\Plaintitle{Restricted mean time lost for survival
  and competing risks data using mets in R}
\Shorttitle{RMST using \pkg{mets} in \proglang{R}}
\Plainkeywords{Restricted Mean Survival (RMST), Restricted Mean Time Lost
  (RMTL), Inverse Probability of Censoring (IPCW), Influence Function, R}

\Address{
  Thomas Scheike\\
  Section of Biostatistics\\
  Institute of Public Health\\
  University of Copenhagen\\
  Øster Farimagsgade 5B, 1014 Copenhagen \\
  Denmark \\
  E-mail: \email{ts@biostat.ku.dk}\\
  ~\\
  Klaus Kähler Holst\\
  Novo Nordisk A/S\\
  Vandtårnsvej 112, 2860 Søborg\\
  Denmark\\
  E-mail: \email{kkzh@novonordisk.com}
}

\Abstract{This paper introduces software implemented in the \pkg{mets} R-package for calculating
non-parametric and regression estimates of Restricted Mean Survival Time (RMST)
and Restricted Mean Time Lost (RMTL), including RMTL due to
specific causes. A unique feature is the ability to compute the non-parametric
estimates of RMST and RMTL, as well as their standard errors, for all time
horizons simultaneously.
Regression modeling in mets is based on Inverse Probability of Censoring
Weighting (IPCW) methods. The package implements different versions of IPCW
adjusted estimating equations. 
A critical
technical contribution is the provision of influence functions for all models,
which enables the computation of standard errors and allows the estimates to be
used as building blocks for more complex statistics, such as the while-alive
estimate in recurrent events settings. To expand capabilities in causal
inference, the \pkg{mets} package also implements methods for standardization
estimates (G-computation) and the estimation of Average Treatment Effects (ATE)
for both RMST and RMTL in the competing risks setting. Importantly, the
computations scale linearly with the number of observations, making the software
efficient for use with large datasets.
}

\begin{document}

\section[Restricted mean time lost]{Restricted mean time lost\label{sec:intro}}

Restricted mean survival (RMST) has become increasingly popular when analysing
time to event data, due to its direct interpretation on the time-scale, giving
the expected number of years alive up to a fixed time horizon. When comparing,
for example, two treatment arms in a medical setting, the effect of the
treatment can be expressed as the number of years lost up to the time horizon
when not on active treatment, see, for
example~\cite{Perego2020,Couchoud2017,AHern2016,Uno2014,Hasegawa2020}. This has
considerable appeal compared to a more traditional comparison of treatments
using the proportional hazards model \citep{cox1972}.

The restricted mean survival, or equivalently, the restricted mean time lost
(RMTL), calculated as the time horizon minus the RMST, similarly provides a
useful summary. Also, in the presence of competing risks, there is interest in
how the restricted time lost can be decomposed into different causes,
see~\cite{Andersen2013} and also~\cite{lyu2020use,Wu2022}.

The RMST can be estimated non-parametrically by the integral of the Kaplan-Meier estimator, and the
RMTL due to different causes can be estimated as the integral of the Aalen-Johansen cumulative incidence
estimator, see~\cite{Andersen2013}, and are thus easy to obtain in practice.

Regression for the restricted mean survival time has been developed and
is based on inverse probability of censoring weighting (IPCW) methods;
see, for example~\cite{Zhao2016,sun2009class,sun2012mean,tian2014},
or by the pseudo-value approach~\cite{andersen-klein-rosthoej-2003,Andersen2010,andersen-ravn-2023,Overgaard2019}.
These methods are well understood and work well in practice.
Similarly, the RMTL for different causes has also been extended to the regression setting in~\cite{Andersen2013,Conner2021}.

We return to the many packages that are dedicated to various aspects of
dealing with RMST in \proglang{R} below, which is the key focus for us.
For completeness we also  mention 
that RMST analysis is supported in \proglang{SAS} via the built-in 
\pkg{PROC RMSTREG} (SAS/STAT 9.4M5+), which fits linear or log-linear models 
using pseudovalue or IPCW methods. 
In \proglang{Stata}, the user-written 
\pkg{strmst2} command calculates RMST and estimates covariate effects,
while there is also support RMST for 
post-estimation following flexible parametric models 
\pkg{stpm2}.

We provide software for computing the non-parametric estimates of RMST and
RMTL in the competing risks setting  \citep{mets-package}.
The RMST is implemented in several other R-packages, for example, 
the \pkg{survival} \citep{survival-package},  
\pkg{eventglm} package  \citep{rmeanglm,SachsGabriel_eventglm},
and the \pkg{survRM2} package  \citep{survRM2-package}.
In our package, a nice
feature is the RMST is computed for all time-horizons at once. In addition, we provide the influence functions for the
estimates which makes it possible to use the RMST as a building block in other settings, and we here give one such example computing
a while-alive estimate in the recurrent events setting that is defined as the ratio of the average number of events for recurrent events and the RMST, see~\cite{mao2023,schmidli2021estimands}, that is implemented in the \pkg{WA}-package \cite{wa-r}.
We compute the estimate of the RMTL and its standard error using the formula of~\cite{Wu2022,bajorunaite2007two} across all time-horizons at once.
To the best of our knowledge, this is not available in other packages. The
\pkg{lillies} package \citep{lillies-package} also computes the estimates but standard errors are based on
bootstrap. Furthermore, we also provide influence functions for the RMTL due to a specific cause, and estimates can therefore be combined with the other estimands.

In the \pkg{mets} package \citep{mets-package} we have also developed regression modelling of the RMST and RMTL based on different versions of IPCW adjusted estimating equations. All regression models
have standard errors computed based on their influence functions that are returned for all models.
RMST regression has also been implemented nicely in the
\pkg{survRM2}-packages and our implementation provide a supplement to theirs and has the advantage that we consider different IPCW estimating equations and
return influence functions. Further, our software can also be used in the competing risks setting with the RTML due to a specific cause. An important feature is
that the censoring adjustment weights can be allowed to depend on strata, and
that this can be achieved without increasing the computational complexity.
All regression models can be used for predictions with computation of standard errors.

We have also implemented standardisation estimates (G-computation) and the estimation of average treatment effects in the causal
setting for both RMST and RMTL for competing risks. For these estimates, standard errors are again computed based on the derived
influence functions. These features are not implemented elsewhere to the best of our knowledge.  Finally, we stress that our computations scale linearly
in the number of observations and therefore can be used for large data, unlike the \pkg{survRM2} and \pkg{lillies} packages.

There are  several other nice recent developments 
to support inference based on restricted mean survival time (RMST) and related
measures. For example, the \pkg{survRM2perm} \citep{survRM2perm} 
package implements
permutation-based RMST comparisons, addressing small-sample limitations of
asymptotic tests and handling censoring issues. The package \pkg{rmt}
\citep{rmt}
provides inferential and graphical tools for comparing treatment groups using the
restricted mean time in favor of treatment. Adaptive procedures for
selecting truncation time points are available in \pkg{survRM2adapt}
\citep{survRM2adapt}.
More recently, \pkg{GFDrmst} \citep{GFDrmst} extends RMST-based inference to general
factorial designs, offering asymptotic, bootstrap, and permutation tests along
with confidence intervals. Similarly, \pkg{GFDrmtl} \citep{GFDrmtl} 
focuses on restricted
mean time lost (RMTL) in competing risks settings and provides multiple testing
procedures. Together, these packages highlight the growing interest in
RMST- and RMTL-based analysis.

The remainder of this paper is organised as follows. In Section~\ref{sec:rmst-rmtl} we describe
the mathematical notation and setup, and we introduce non-parametric estimators
for the RMST and RMTL. Throughout the paper, we illustrate the methodology with
concrete examples using the \pkg{mets} package in \proglang{R}. In Section~\ref{sec:regression} we consider IPCW-based regression models
for both these quantities, and also describe the extension to estimation of
Average Treatment Effects (ATE). A real-world example is presented in
Section~\ref{sec:examples}, where we analyse a randomised clinical trial
(HF-ACTION) investigating the impact of exercise training on heart-failure endpoints
and here we estimate both the ATE for the RMST and also consider a while-alive
estimand. 
In  Section~\ref{sec:cif-reg} we briefly make the observation that the 
machinery outlined in the paper can be used also for competing risks regression. 
Finally, a discussion and conclusion are presented in Section~\ref{sec:discussion}.

\section{RMST and RMTL for right censored data} \label{sec:rmst-rmtl}

Let $D$ be the time-to-event with cause of death $\eta \in \{ 1,2\}$, that is
observed subject to right censoring. Specifically, because of a censoring time
$C$, we observe ${T}=D\wedge C$, the right censored time to event, $\Delta= I(D \leq C)$,
the event (or non-censoring) indicator and the observed cause of death ${\epsilon}=\Delta\eta$.
Here we use the notation $a \wedge b$ to denote the minimum of $a$ and $b$.
In addition to $(T,\Delta , \epsilon)$, we also observe baseline covariates $X$.
We will assume that we have independent right-censoring, thus assuming
that $C$ and $D$ are independent given $X$. In addition we will for simplicity
of computations assume that the censoring distribution given $X$ depends only on $X$ though
a finite set of strata $L(X)$, and write its distribution as
$G_c(t,L(X)) = \Prob(C > t | L(X))$. We will need that $G_c (t,L(X)) > \epsilon > 0 $ for
all strata $L(X)$ to guarantee that we observations to identify the estimands of interest.
We let $S(t)=\Prob(D \geq t)$ denote the survival distribution of $D$.  We make the smoothness assumptions
that the  related cause specific hazards are $\alpha_j(t;X)$ for $j=1,2$ and that of $C$ is $\lambda_c(t;L(X))$.
Define further the observed data censoring 
martingale
$M_c(t)=I(D\leq t, \Delta=0)-\int_0^t I(T\geq u) \lambda_c(u|L(X))du$,
for the censoring process with hazard function $\lambda_c(t|L(X))$.
We also write $\Lambda_c(t,L(x))= \int_0^t \lambda_c(u|L(x)) du$ for the cumulative hazard.

We assume that we have available observations
$({T}_i,\Delta_i,{\epsilon_i},{X}_i)$, $i=1,\dots,n$, that are i.i.d. copies of $(T,\Delta,\eta,{X})$.

We shall now be interested in regression and non-parametric modelling of outcomes
such as the restricted mean survival time (RMST) $Z(\tau) = (D \wedge \tau)$ or
the restricted mean time lost (RMTL) $R(\tau)=\tau - (D \wedge \tau)$ for some
specific time horizon $\tau$.
Further, in the competing risks setting we
shall also consider the RMTL due to cause $1$, say, $R_1(\tau) = I(\eta=1) ( \tau- D \wedge \tau)$, the time-lost due to cause $1$.
We note that we have that $R(\tau)=R_1(\tau)+R_2(\tau)$.
The specific time horizon $t$ is typically specified by subject matter experts
and may by $1$ or $5$ years in many clinical trials.

In the case without covariates, or within strata,
a nonparametric maximum likelihood estimate of the RMST can be obtained by integrating the Kaplan-Meier estimator, since
$\E(D \wedge \tau) = \int_0^\tau S(s)ds$ that therefore can be estimated by
\begin{eqnarray*}
\hat \psi    = \int_0^\tau \hat S(s) ds
\end{eqnarray*}
where $\hat S(s)$ is the Kaplan-Meier estimator of $S(s)$. The asymptotic properties of $\hat \psi$ were given
in for example \cite{abgk} Example I.V.3.8. and $\hat \psi$ is asymptotically normal with mean $\int_0^\tau S(s)ds$ and
a variance that may be estimated by
\begin{eqnarray*}
	 \int_0^\tau \left\{ \int_s^\tau \hat S(u) ds\right\}^2  \frac{1}{Y(s) (Y(s)-1)} dN_1(s) + dN_2(s),
\end{eqnarray*}
where $N_j(s)$ counts
the number of events of cause $j=1,2$ and $Y(s)$ the total number of subjects under risk just prior to time $s$.

To illustrate the methods we will be using the \code{bmt} data that briefly is
data from myelodysplasia patients treated with HLA-identical sibling bone marrow transplantation. 
The study considered the competing risks responses: treatment-related mortality (TRM, cause=1), defined as death in complete remission,
and relapse (cause=2); see \cite{blood}.
The study comprised 408 patients with complete information of platelet counts at transplantation.
The following three risk factors were considered in the regression models:
platelet counts, either $> 100\times 10^9/\textrm{L}$ (31 \%) or
$\leq 100\times 10^9 /\textrm{L}$ (69 \%);
age, a continuous variable, standardized and centered at the mean of 35-years-old and ranging from 2 to 64-years-old;
and graft-versus-host disease prophylaxis, either T-cell depletion (13\%) or no T-cell depletion (87\%).

\begin{Schunk}
\begin{Sinput}
R> library("mets")
R> data("bmt", package = "mets")
R> set.seed(1)
R> bmt$time <- bmt$time + runif(nrow(bmt)) * 0.001 # break ties
R> 
R> out1 <- phreg(Surv(time, cause != 0) ~ strata(tcell, platelet), data = bmt)
R> rm1 <- resmean_phreg(out1, times = 30)
R> summary(rm1, contrast = rbind(c(1, -1, 0, 0))) 
\end{Sinput}
\begin{Soutput}
$estimates
                    strata times    rmean  se.rmean    lower    upper
tcell=0, platelet=0      0    30 13.60294 0.8315416 12.06700 15.33439
tcell=0, platelet=1      1    30 18.90126 1.2693308 16.57019 21.56026
tcell=1, platelet=0      2    30 16.19122 2.4006108 12.10808 21.65130
tcell=1, platelet=1      3    30 17.76608 2.4422055 13.57005 23.25956
                    years.lost
tcell=0, platelet=0   16.39706
tcell=0, platelet=1   11.09874
tcell=1, platelet=0   13.80878
tcell=1, platelet=1   12.23392

$test
                          Estimate Std.Err   2.5%  97.5%   P-value
[tcell=0, platelet=0]....   -5.298   1.517 -8.272 -2.324 0.0004802
------------------------------------------------------------
Null Hypothesis: 
  [tcell=0, platelet=0] - [tcell=0, platelet=1] = 0 
 
chisq = 12.1912, df = 1, p-value = 0.0004802
\end{Soutput}
\end{Schunk}

\begin{figure}[t!]
\centering
\begin{Schunk}

\input{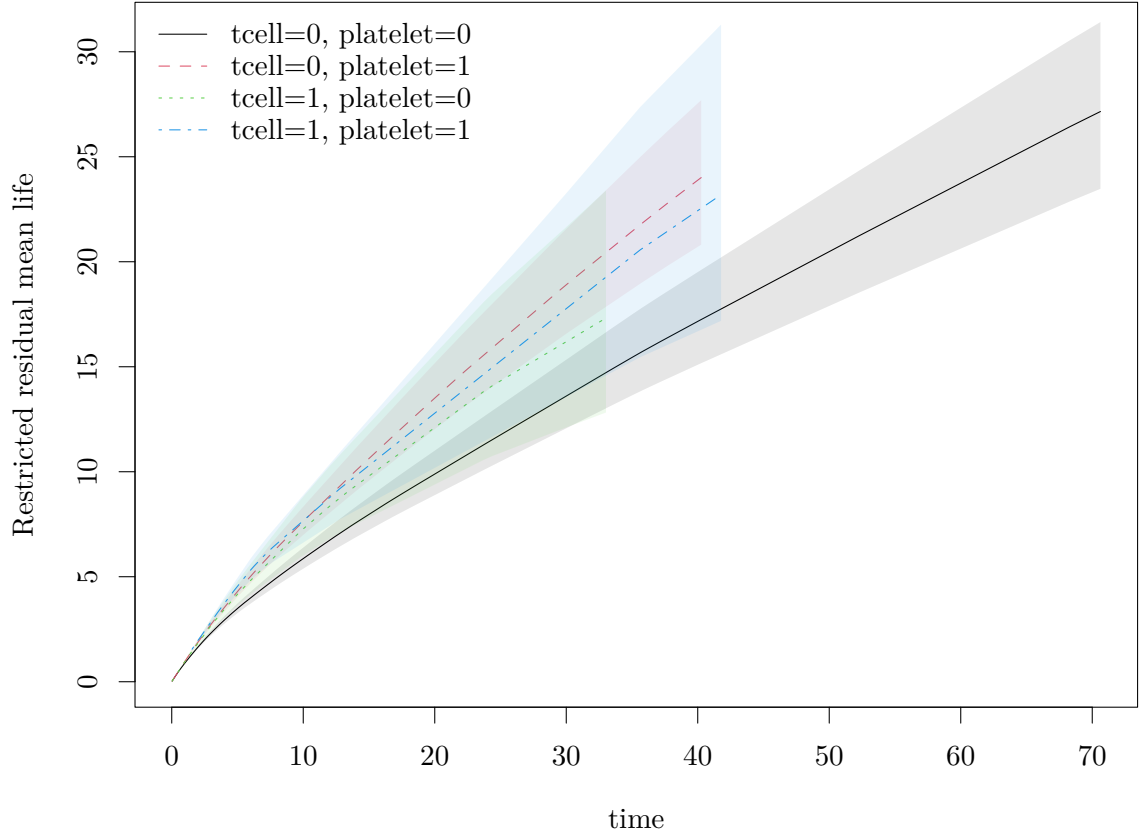}\end{Schunk}
\caption{\label{fig:rmst} RMST for all time-horizons with 95 \% confidence intervals.}
\end{figure}

We observe that strata 0, the group with no T-cell depletion and low 
platelet count, has an estimated RMST at 
$13.6$ with 95 \% confidence interval ($12.1,15.3$), or in other words looses
on average $16.4$ years up to 30 years, 
and with a difference in  RMST between  strata 0 and strata 1 
of $-5.3$ ($-8.7,-2.3$) years.
Figure~\ref{fig:rmst} shows the RMST for different time-horizons.

As pointed out by \cite{Andersen2013} the total years lost $\tau - \int_0^\tau S(s) ds  = \int_0^\tau (1- S(s)) ds$ 
and therefore the total years lost can be decomposed in the the years lost due to the different causes since 
\begin{eqnarray*}
	\tau - \int_0^\tau S(s) ds   = \int_0^\tau F_1(s) ds + \int_0^\tau F_2(s) ds.
\end{eqnarray*}
This decomposition holds for any number of causes.
The non-parametric maximum likelihood estimator is therefore given as 
\begin{eqnarray*}
 \hat \psi_1  = \int_0^\tau \hat F_1(s) ds,
\end{eqnarray*}
where $\hat F_1(\tau)= \int_0^\tau [ \hat S(s-)/Y(s)] dN_1(s)$ is the Aalen-Johansen product limit estimator.
The asymptotic properties of $\hat \psi_1$ were given in 
\cite{lin1997nonparametric,Wu2022,bajorunaite2007two}
and $\hat \psi_1$ is asymptotically normal with mean $\int_0^\tau F_1(s)ds$ and a variance that may be estimated by 
\begin{eqnarray*}
 \int_0^\tau [ (\tau - s) ( 1-\hat F_2(s)) - \int_s^\tau \hat F_1(u) du ]^2 \frac{1}{Y(s) (Y(s)-1)} dN_1(s)  \\
 +  \int_0^\tau [ (\tau - s) \hat F_1(s)     -  \int_s^\tau \hat F_1(u) du ]^2 \frac{1}{Y(s) (Y(s)-1)} dN_2(s).
\end{eqnarray*}

\begin{Schunk}
\begin{Sinput}
R> rmc1 <- cif_yearslost(Event(time,cause) ~ strata(tcell, platelet), 
+           data = bmt, times = 30)
R> summary(rmc1, contrast = rbind(c(1, -1, 0 ,0)))
\end{Sinput}
\begin{Soutput}
$testintF_1
            Estimate Std.Err  2.5% 97.5%   P-value
[p1] - [p2]    5.221    1.45 2.379 8.063 0.0003173
------------------------------------------------------------
Null Hypothesis: 
  [p1] - [p2] = 0 
 
chisq = 12.9653, df = 1, p-value = 0.0003173

$testintF_2
            Estimate Std.Err  2.5% 97.5% P-value
[p1] - [p2]  0.07738   1.095 -2.07 2.224  0.9437
------------------------------------------------------------
Null Hypothesis: 
  [p1] - [p2] = 0 
 
chisq = 0.005, df = 1, p-value = 0.9437

$estimate
$estimate$intF_1
                    strata times    intF_1 se.intF_1 lower_intF_1
tcell=0, platelet=0      0    30 12.105128 0.8508099    10.547332
tcell=0, platelet=1      1    30  6.884196 1.1741037     4.928108
tcell=1, platelet=0      2    30  7.260708 2.3532722     3.846764
tcell=1, platelet=1      3    30  5.780382 2.0925012     2.843290
                    upper_intF_1
tcell=0, platelet=0    13.893004
tcell=0, platelet=1     9.616702
tcell=1, platelet=0    13.704474
tcell=1, platelet=1    11.751463

$estimate$intF_2
                    strata times   intF_2 se.intF_2 lower_intF_2
tcell=0, platelet=0      0    30 4.291929 0.6161439     3.239329
tcell=0, platelet=1      1    30 4.214544 0.9057029     2.765845
tcell=1, platelet=0      2    30 6.548072 1.9703413     3.630642
tcell=1, platelet=1      3    30 6.453541 2.0815255     3.429663
                    upper_intF_2
tcell=0, platelet=0     5.686564
tcell=0, platelet=1     6.422044
tcell=1, platelet=0    11.809824
tcell=1, platelet=1    12.143522


$total.years.lost
[1] 16.39706 11.09874 13.80878 12.23392
\end{Soutput}
\end{Schunk}

\begin{figure}[t!]
\centering
\begin{Schunk}

\input{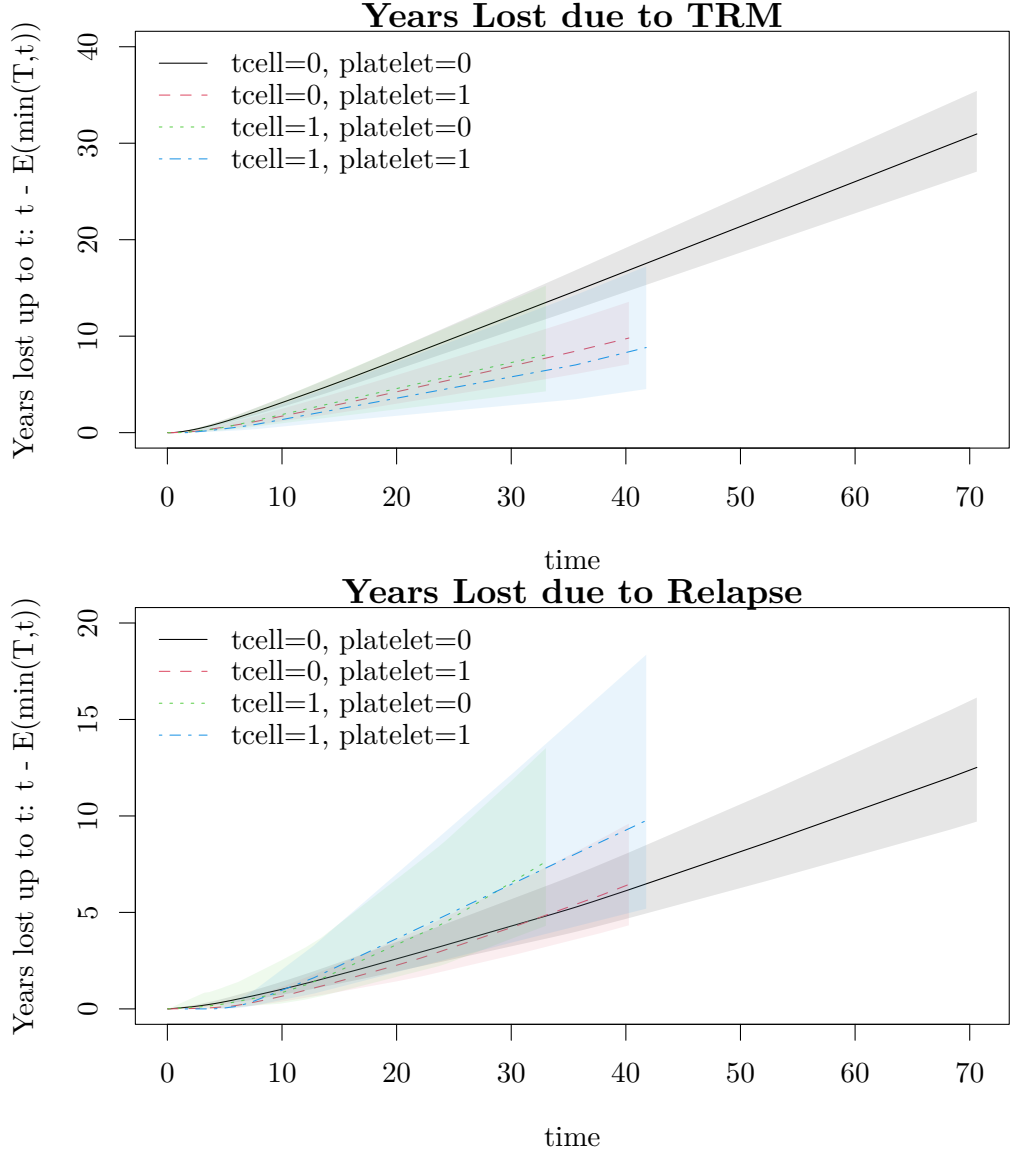}\end{Schunk}
\caption{\label{fig:rmtl-12} RMTL for TRM and relapse for all time-horizons with 95 \% confidence intervals.}
\end{figure}

We observe that strata 0 looses $12.1$ ($10.5,13.9$) years to cause 1, and
$4.2$ ($3.2,5.7$) years to cause 2, and when comparing the 
two strata, we conclude that strata 0  thus looses 
$5.2$ ($2.4,8.1$) years more than strata 1.
Figure~\ref{fig:rmtl-12} shows the years lost for different time-horizons for TRM and relapse.

\section{IPCW-based Regression}\label{sec:regression}

We now consider how to do inverse probability of censoring weighted regression for the RMST, RMTL or the RMTL due to a specific cause.
If any of the outcomes $Z(\tau)$, $R(\tau)$ or $R_j(\tau)$ for $j=1,2$ were observed we could now consider different regression models by solving
the estimating equation
\begin{equation}
	U(\beta) =  \frac{1}{n}\sum_{i=1}^n X_i  \{ O_i(\tau) - F(\tau,X_i^T \beta)) \} =  \bf{0} \label{eq:uncensored}
\end{equation}
where $O(\tau)$ is any of the outcomes, $F(\cdot)$ is a smooth nice link function such as the exponential or the indentity,
$X_i$ is the covariates and $\beta$ a set of regression coefficients.
We denote the solution to the score equation as $\beta^\ast$, that is
$E\{U(\beta^\ast)\}=0$.

To deal with the right censoring it has been suggested to use inverse probability of censoring weights to modify this
estimating equation. Define censoring weights as $\hat W(\tau) = W(\tau,\hat G_c) = I(D \wedge \tau \leq C)/ \hat G_c(D \wedge \tau, L(X))$, where
$\hat G_c(\cdot,L(X))$ are stratified Kaplan-Meier estimators. We define similarly $W(\tau)=W(\tau,G_c)$.
Importantly, if $W(\tau) > 0$ then we can compute the full data outcome $O(\tau)$.

We consider the outcome weighted IPCW approach \citep{Scheike2008a,Blanche2022} that then yields the IPCW adjusted estimating equation
\begin{equation}
	U^{\textrm{IPCW}}(\beta) =  \frac{1}{n}\sum_{i=1}^n X_i \{ \hat W_i(\tau) O_i(\tau) - F(\tau,X_i^T \beta)) \} =  \bf{0} \label{eq:censored}.
\end{equation}
Our estimator is found as the solution to $U^{\textrm{IPCW}}(\hat \beta) \equiv 0$, and under standard regularity conditions
it can be shown that $\sqrt{n} (  \hat \beta - \beta^\ast)$ is asymptotically linear with influence functions
of the score
\begin{equation}
X_i \{ W_i(\tau) O_i(\tau) - F(\tau, X_i^T \beta^\ast)) \} + \int_0^\tau \E( X_i O_i(\tau) | D \geq s, L(X_i)) \frac{dM_{c,i}(s)}{G_c(s,L(X_i))} \label{if:censored}.
\end{equation}
This can also be written as the full data estimating equation and a censoring term
\begin{equation}
     X_i \{ Z_i(\tau) - F(\tau, X_i^T \beta^\ast)) \} - \int_0^\tau  \{ O_i(\tau) - \E( X_i O_i(\tau) | D \geq s, L(X_i))  \}  \frac{dM_{c,i}(s)}{G_c(s,L(X_i))},
\end{equation}
the term $\E[ X_i Z_i(\tau) | D \geq s, L(X_i)]$ is due to the fact that we estimated the censoring weights \citep{Blanche2022}.
We denote this estimator as the estimator of \code{type="I"} below.

In fact, from the theory of semi-parametric inference we know that all estimators can be written as some full data estimating equation and a
censoring augmentation \citep{tsiatis2006semiparametric,robins-rotnitzky:1992}.

We also estimate $\beta^\ast$ by considering the augmented IPCW estimating equation that have the form
\begin{eqnarray*}
	U^{\textrm{IPCW}}(\beta)  +
	\frac{1}{n}\sum_{i=1}^n \int_0^\tau X_i \widehat \E(O_i(\tau) | D \geq s, L(X_i)) \frac{d \hat M_{c,i}(s)}{\hat G_c(s,L(X_i))}
\end{eqnarray*}
with $\hat M_{c,i}(s)$ the standard estimated censoring martingale and with the IPCW estimator for the conditional mean among survivors
\begin{eqnarray*}
\widehat \E[O_i(\tau) | D \geq s, L(X_i)=k] =  \hat G_c(s,k) \frac{ \sum_i O_i(\tau) \hat W_i(\tau) I(T_i \geq s) I(L(X_i)=k) }{\sum_i  I(L(X_i)=k) I(T_i \geq s)}
\end{eqnarray*}
This is a simple stratified estimator of the response among survivors, and is equivalent to the direct Kaplan-Meier based estimator
$\int_s^\tau \hat S(u,k) du /\hat S(s,k)$  for the RMST outcome, and the stratified Aalen-Johansen estimator
$\int_s^\tau \hat F_1(u,k) du/\hat S(s,k)$ for the RMTL outcomes, see for example \cite{satten2001kaplan,Cortese2017,geskus-2010cause}.
This estimator is denoted as the  \code{type="II"} estimator and is the default estimator.

It can be shown that the influence function of this score function
is
\begin{eqnarray*}
\psi  =  X_i \{ W_i(\tau) O_i(\tau) - F(\tau, X_i^T \beta^\ast)) \} +
\int_0^t \E( X_i O_i(\tau) | D \geq s, L(X_i)) \frac{dM_{c,i}(s)}{G_c(s,L(X_i))} \nonumber  \\
 + \int_0^t [ X_i \E(O_i(\tau) | D \geq s, L(X_i)) - \E( X_i | D \geq s, L(X_i))  \E(O_i(\tau) | D \geq s, L(X_i)) ] \frac{d M_{c,i}(s)}{G_c(s,L(X_i))}  \label{ifpse:censored}.
\end{eqnarray*}
This particularly augmented IPCW estimator will thus have the same asymptotics as the pseudo-value estimator, see
\cite{overgaardIPCW2024,Overgaard2019,parner-andersen-overgaard-2023}.

For both types of regression estimators we can estimate the variance of the
regression coefficients $\hat \beta$ by the sandwich formula using the influence
functions
\begin{eqnarray*}
   \hat \Psi^{-1} \sum_i \hat \psi_i^{\otimes 2} \hat \Psi^{-1}
\end{eqnarray*}
with $\hat \Psi = D_\beta U^{\textrm{IPCW}} ( \hat \beta)$, the derivative of the estimating equation with respect to $\beta$ evaluated at $\hat \beta$.

We illustrate how the estimating equations of both types can be
solved calling the function \code{mets::rmstIPCW} below.


We finally remark that the IPCW adjusted estimating equations of
\cite{tian2014} that are implemented in the \pkg{RMST2}-packages can also be obtained from the \pkg{mets}-package using the
\code{mets::logitIPCW} function specifying outcome and model. Here the estimating equations are
\begin{equation}
	U^{\textrm{GLM}}(\beta) =  \frac{1}{n}\sum_{i=1}^n X_i \hat W_i(\tau) \{ O_i(\tau) - F(\tau,X_i^T \beta)) \} =  \bf{0} \label{eq:glm},
\end{equation}
that we denote as IPCW-GLM estimating equations.
Which IPCW method that is the best depends on the setting \citep{Blanche2022,overgaardIPCW2024}, but in our experience the
\code{mets::rmstIPCW} with \texttt{type="II"} (the default) performs well in many settings.

\subsection{Equivalence with non-parametric estimates}

In the case where $X=L(X)$ so that $X$ is factor with levels for each strata of $L(X)$ and we estimate the censorings weights using the
stratified Kaplan-Meier, $\hat G_c(\cdot,L(X)$ then the predictions based on the IPCW regression model for
the RMST, the RMTL and the RMTL due a specific cause all outcomes use a full censoring model are equivalent to the
non-parametric estimates based on the Kaplan-Meier or the Aalen-Johansen estimators. This holds if there are no ties in the survival times and
otherwise various technical modifications are needed, see \cite{satten2001kaplan,Cortese2017,geskus-2010cause}.


In this case the \code{type="II"} augmentation is $0$ and consequently, \code{type="I"} and
\code{type="II"} estimates are equivalent. When the status is binary
(survival setting, \code{cause!=0} in the below call) the default is
to do the RMST  regression, and when it is has more than two values the 
default is to do RMTL regression (competing risks setting). 
These defaults can be overruled by specifying the outcome of the model

\begin{Schunk}
\begin{Sinput}
R> bmt$int <- with(bmt, strata(tcell, platelet))
R> out <- rmstIPCW(Event(time, cause!=0) ~ -1 + int,
+                  data = bmt, time = 30,
+                  outcome = "rmst", model="lin",
+                  cens.model = ~ strata(platelet, tcell))
R> estimate(out)
\end{Sinput}
\begin{Soutput}
                       Estimate Std.Err  2.5% 97.5%   P-value
inttcell=0, platelet=0    13.60  0.8316 11.97 15.23 3.825e-60
inttcell=0, platelet=1    18.90  1.2696 16.41 21.39 4.001e-50
inttcell=1, platelet=0    16.19  2.4061 11.48 20.91 1.704e-11
inttcell=1, platelet=1    17.77  2.4536 12.96 22.58 4.464e-13
\end{Soutput}
\begin{Sinput}
R> estimate(rm1) # estimates based on KM
\end{Sinput}
\begin{Soutput}
                    Estimate Std.Err  2.5% 97.5%   P-value
tcell=0, platelet=0    13.60  0.8315 11.97 15.23 3.771e-60
tcell=0, platelet=1    18.90  1.2693 16.41 21.39 3.786e-50
tcell=1, platelet=0    16.19  2.4006 11.49 20.90 1.534e-11
tcell=1, platelet=1    17.77  2.4422 12.98 22.55 3.474e-13
\end{Soutput}
\begin{Sinput}
R> head(IC(out)) # estimated influence function
\end{Sinput}
\begin{Soutput}
          [,1] [,2] [,3] [,4]
[1,] -21.80178    0    0    0
[2,] -21.80161    0    0    0
[3,] -21.80129    0    0    0
[4,] -21.80075    0    0    0
[5,] -21.80188    0    0    0
[6,] -21.80076    0    0    0
\end{Soutput}
\end{Schunk}

estimates are equivalent to those based on the Kaplan-Meier (\code{rm1})
when  there are no ties, as illustrated, 
but the standard errors are estimated differently. 
An advantage of the IPCW estimator is that the influence functions are available.

Now, looking at the RMTL due to the cause=1, we use the IPCW regression to get

\begin{Schunk}
\begin{Sinput}
R> out1 <- rmstIPCW(Event(time, cause) ~ -1 + int,
+                   data = bmt, time = 30, cause = 1,
+                   cens.model = ~ strata(platelet, tcell), model = "lin")
R> estimate(out1)
\end{Sinput}
\begin{Soutput}
                       Estimate Std.Err   2.5%  97.5%   P-value
inttcell=0, platelet=0   12.105  0.8508 10.438 13.773 6.162e-46
inttcell=0, platelet=1    6.884  1.1741  4.583  9.185 4.536e-09
inttcell=1, platelet=0    7.261  2.3533  2.648 11.873 2.033e-03
inttcell=1, platelet=1    5.780  2.0925  1.679  9.882 5.737e-03
\end{Soutput}
\begin{Sinput}
R> estimate(rmc1) # estimates based on AJ
\end{Sinput}
\begin{Soutput}
                    Estimate Std.Err   2.5%  97.5%   P-value
tcell=0, platelet=0   12.105  0.8508 10.438 13.773 6.161e-46
tcell=0, platelet=1    6.884  1.1741  4.583  9.185 4.536e-09
tcell=1, platelet=0    7.261  2.3533  2.648 11.873 2.033e-03
tcell=1, platelet=1    5.780  2.0925  1.679  9.882 5.737e-03
\end{Soutput}
\begin{Sinput}
R> head(IC(out)) # estimated influence function
\end{Sinput}
\begin{Soutput}
          [,1] [,2] [,3] [,4]
[1,] -21.80178    0    0    0
[2,] -21.80161    0    0    0
[3,] -21.80129    0    0    0
[4,] -21.80075    0    0    0
[5,] -21.80188    0    0    0
[6,] -21.80076    0    0    0
\end{Soutput}
\end{Schunk}

Again, we note that the IPCW estimates are equivalent to those based on
the integrated Aalen-Johansen estimator (\code{rmc1}). Standard errors are still consistent
but based on an alternative representation from the influence functions in
the regression setting. 

\subsection{General IPCW regression}

We now do regression modeling for the total RMST using the exponential link function.

\begin{Schunk}
\begin{Sinput}
R> outrmtl <- rmstIPCW(Event(time, cause !=0) ~ tcell + platelet + age,
+                      data = bmt, time = 30,
+                      cens.model = ~ strata(platelet, tcell),
+                      model="exp")
R> summary(outrmtl)
\end{Sinput}
\begin{Soutput}
   n events
 408    231

 408 clusters
coeffients:
             Estimate   Std.Err      2.5%     97.5% P-value
(Intercept)  2.610518  0.057566  2.497692  2.723345  0.0000
tcell        0.148799  0.119047 -0.084529  0.382128  0.2113
platelet     0.243180  0.084230  0.078092  0.408269  0.0039
age         -0.172771  0.037858 -0.246972 -0.098570  0.0000

exp(coeffients):
            Estimate     2.5%   97.5%
(Intercept) 13.60610 12.15440 15.2312
tcell        1.16044  0.91895  1.4654
platelet     1.27530  1.08122  1.5042
age          0.84133  0.78116  0.9061
\end{Soutput}
\begin{Sinput}
R> head(IC(outrmtl)) # estimated influence function
\end{Sinput}
\begin{Soutput}
          [,1]      [,2]      [,3]        [,4]
[1,] -1.497966 0.8831440 1.2055215 -0.29475646
[2,] -1.443072 1.0292216 1.0179123 -0.60603175
[3,] -1.342049 1.2333471 0.7246608 -1.06209036
[4,] -1.345753 1.2270529 0.7344566 -1.04751688
[5,] -1.529026 0.7875234 1.3220973 -0.09522617
[6,] -1.303522 1.2922810 0.6280262 -1.20187974
\end{Soutput}
\end{Schunk}

and regression for the RMTL due to cause 1 with exponential link

\begin{Schunk}
\begin{Sinput}
R> outrmtl1 <- rmstIPCW(Event(time, cause) ~ tcell + platelet +age,
+                       data = bmt, time = 30, cause = 1,
+                       cens.model = ~strata(platelet, tcell),
+                       model = "exp")
R> summary(outrmtl1)
\end{Sinput}
\begin{Soutput}
   n events
 408    154

 408 clusters
coeffients:
              Estimate    Std.Err       2.5%      97.5% P-value
(Intercept)  2.4461311  0.0729852  2.3030827  2.5891795  0.0000
tcell       -0.4961306  0.2513854 -0.9888370 -0.0034242  0.0484
platelet    -0.4547583  0.1675116 -0.7830750 -0.1264415  0.0066
age          0.2748792  0.0654241  0.1466504  0.4031080  0.0000

exp(coeffients):
            Estimate     2.5%   97.5%
(Intercept) 11.54360 10.00498 13.3188
tcell        0.60888  0.37201  0.9966
platelet     0.63460  0.45700  0.8812
age          1.31637  1.15795  1.4965
\end{Soutput}
\begin{Sinput}
R> head(IC(outrmtl1)) # estimated influence function
\end{Sinput}
\begin{Soutput}
          [,1]      [,2]     [,3]       [,4]
[1,] -1.614005 0.9851675 1.440903  0.0967899
[2,] -1.628700 1.3801272 1.469911 -0.6198790
[3,] -1.593043 2.3426233 1.478592 -2.4517463
[4,] -1.596182 2.3003816 1.479577 -2.3694692
[5,] -1.597681 0.7935567 1.418853  0.4334913
[6,] -1.550802 2.8111126 1.461228 -3.3731255
\end{Soutput}
\end{Schunk}

Regression effects are interpreted as giving log-proportional effects, such
that for example T-cell depletion (\code{tcell}=1) leads to an RMST that
is 16 \% higher, $1.16 (0.92, 1.47)$, all other things fixed.  Similarly, 
the RMTL due
to TRM is 40 \% lower, $0.6 (0.4,1.0)$, for those who are T-cell depleted.

We can predict the RMST or RMTL based on the regression models using the a
predict function and can ask to get the estimated influence functions to use for
further work or when combining predictions

\begin{Schunk}
\begin{Sinput}
R> nd <- data.frame(tcell = c(0, 1, 0), platelet=c(1, 0, 1), age=c(2, 0, -1))
R> pnd <- predict(outrmtl, nd)
R> pnd
\end{Sinput}
\begin{Soutput}
      pred       se     lower    upper
1 12.28227 1.504140  9.334205 15.23033
2 15.78907 1.857989 12.147477 19.43066
3 20.62429 1.346219 17.985751 23.26283
\end{Soutput}
\end{Schunk}

For example,  for a patient with T-cell depletion and high platelet count, and
an age that is 2 sd's larger than the mean age the predicted RMST is
$12.3 (9.3,15,2)$ years.

%

Finally, as mentioned earlier the \cite{tian2014} estimating equations can be solved using

\begin{Schunk}
\begin{Sinput}
R> outglm <- logitIPCW(Event(time, cause != 0) ~ tcell + platelet + age,
+                      data = bmt, time=30,
+                      outcome = "rmst", model = "lin",
+                      cens.model = ~ strata(platelet, tcell))
R> summary(outglm)
\end{Sinput}
\begin{Soutput}
   n   events
 408 5582.529

 408 clusters
coeffients:
            Estimate  Std.Err     2.5%    97.5% P-value
(Intercept) 13.69026  0.84910 12.02606 15.35445  0.0000
tcell        2.57138  2.19529 -1.73130  6.87407  0.2415
platelet     3.96255  1.60785  0.81122  7.11388  0.0137
age         -3.05385  0.66712 -4.36138 -1.74632  0.0000
\end{Soutput}
\begin{Sinput}
R> head(IC(outglm)) # estimated influence function
\end{Sinput}
\begin{Soutput}
          [,1]     [,2]      [,3]       [,4]
[1,] -20.12151 12.60975 16.938047  -2.966200
[2,] -18.21317 14.36573 13.939938  -7.890125
[3,] -14.60749 15.80395  9.161360 -13.879180
[4,] -14.74345 15.79178  9.321716 -13.727517
[5,] -21.19083 11.30796 18.767851   0.353337
[6,] -13.19048 15.73247  7.583590 -15.110110
\end{Soutput}
\end{Schunk}

We note that estimates based on these IPCW estimating equations are
quite similar.

\subsection{Standardization and ATE}

In this Section we demonstrate how the efficient and doubly robust estimator
of the average treatment effect can be computed for RMST or RMTL, based on
observational data under the standard causal assumptions \cite{hernan_robins_2020}.

We need a little bit of notation.
Let $A_i$ denote the treatment indicator and let $X_i$ be additional baseline
covariates. Define the propensity score as
$ \pi_A(X_i) = \Prob(A_i = 1 \mid X_i)$.
Let $ \tilde{O}_i(\tau) = O_i(\tau)\,\hat{W}_i(\tau)$ be the IPCW adjusted
outcome,  where $\hat{W}_i(\tau)$ are 
inverse probability of censoring weights.
We are interested in the mean of, $O^a(\tau)$, the counterfactual outcome 
if all received treatment $a$, $a=0,1$.
Furthermore, define the outcome regression model as
$ m_j(X_i) = \E(O_i \wedge \tau \mid A_i = j, X_i), $
which is estimated using inverse probability of censoring 
weighted (IPCW) regression.
We still assume independent censoring conditional
on $(A, X)$, 
and that the censoring distribution depend
on $Z = (A, X)$ through strata $L(Z)$.
The propensity score model and outcome regression model are
parametrized as a logistic regression model  $\pi(X,\alpha)$ and as 
a regression model such as $m_1(X,\gamma)= \exp(X^T \gamma)$.

The RMST for the treated is $\mu(1)= \E[O^1(\tau)]$ and can be
estimated by solving the estimating equation
\begin{eqnarray}
	U(\hat \pi, \hat m_1, \hat G_c, \mu(1))  = 
    n^{-1} \sum_i  U_i(\hat \pi,\hat m_1,\hat G_c, \mu(1))  = 0  \label{ate-eq}
\end{eqnarray}
with
\begin{eqnarray*}
    U_i(\pi,m_1,G_c,\mu(1))&  = &\left[ \frac{A_i \tilde O_i(\tau)}{\pi_A(X_i)}-\frac{A_i - \pi_A(X_i)}{\pi_A(X_i)} m_1(X_i) - \mu(1)  \right]  \\
                          & = &  \frac{A_i }{\pi_A(X_i)} [ \tilde O_i(\tau) - m_1(X_i) ] +   [ m_1(X_i) - \mu(1)],
\end{eqnarray*}
and similarly for the those not treated $a=0$.
The estimator of the RMST, $\mu(1)$, among the treated is
\begin{eqnarray*}
  \hat \mu(1)   = n^{-1} \sum_i \left[ \frac{A_i \tilde O_i(\tau)}{\hat \pi_A(X_i)}-\frac{A_i - \hat \pi_A(X_i)}{\hat \pi_A(X_i)} \hat m_1(X_i)  \right],
\end{eqnarray*}
and if all models are correct and we use the fully non-parametric stratified Kaplan-Meier this is the efficient estimator,
\cite{tsiatis2006semiparametric,Genetti2025}. The default is to add 
censoring augmentation term to the estimating equation \eqref{ate-eq}
\begin{eqnarray*}
	 &  n^{-1} \sum_i  \int_0^\tau \frac{A_i}{ \hat \pi(X_i)} \hat \E( O(\tau) | T \geq s, L(Z)) \frac{1}{\hat G_c(s,L(Z))} d \hat M_c(s,L(Z))
\end{eqnarray*}
for  \code{typeATE="II"}.

The estimator for the treated has a standard deviation that  can described through its influence function after Taylor-expanding in the
direction of the used models for $\pi(X,\alpha)$, $m_1(X,\gamma)$ and $G_c(s,L(Z))$, that are estimated by logistic regression, IPCW-regression as described in
previous section, and the stratified Kaplan-Meier given strata $L(Z)$.
This leads to the influence function for $\hat \mu(1)$,
{\small
\begin{align*}
  \epsilon_{\mu(1)}  &=   U(\pi,m_1,G_c,\mu(1))  + \E(D_\beta U) \epsilon_{\beta} + \E(D_\gamma U ) \epsilon_{\gamma} \\
	&\quad +\, \int_0^t \E(\frac{A}{\pi(X)} O(t) | T \geq s, L(Z)) \frac{1}{G_c(s,L(Z))} dM_c(s,L(Z)) \\
                     &\quad +\, \int_0^\tau [  \frac{A}{\pi(X)} \E( O(\tau) | T \geq s, L(Z)) \\
  &\quad -\, \E( \frac{A}{\pi(X)} |  T \geq s, L(Z)) \E( O(t) | T \geq s, L(Z))  ] \frac{1}{G_c(s,L(Z))} dM_c(s,L(Z))
\end{align*}
}
where $\epsilon_{\beta}$ is the influence function from the logistic regression, $\epsilon_{\gamma}$ is the influence function from the IPCW-regression for the
outcome model, and the last terms are 
the influence function related to the stratified Kaplan-Meier estimator and
from the \code{typeATE="II"} augmentation term.
The covariance matrix of the ATE estimators for 
the $(\mu(1), \mu(0))$  is thus
given by 
$n^{-1} \sum_i ( \hat \epsilon_{\mu(1)}, \hat \epsilon_{\mu(0)})^{\otimes 2}$.
The standard errors are always valid even under misspecified models.

Consequently, the ATE, $\E[O^1(\tau)]- \E[O^0(\tau)]$, i.e. the difference
in RMST, can be estimated by
\begin{eqnarray*}
 \hat \mu(1) - \hat \mu(0)
\end{eqnarray*}
and its influence function is the sum of the influence functions from above.
Rather than the ATE estimator described above one can also just 
compute the G-estimator using the outcome model
\begin{eqnarray*}
  \hat \mu_G(1) = n^{-1}   \sum_i \hat m_1(X_i,\hat \gamma),
\end{eqnarray*}
thus averaging over the covariate distribution of the data. This 
estimator has influence function
\begin{eqnarray*}
	\epsilon_{\mu_G(1)}  = m_1(X_i,\beta^\ast) - \mu(1)  + \E[D_\gamma \mu_G(1)] \cdot \epsilon_{\gamma}.
\end{eqnarray*}
and the ATE and G-estimator are equivalent when the used outcome model satisfies
that
$\E\{ \frac{A_i }{\pi_A(X_i)} [ \tilde O_i(\tau) - m_1(X_i,\gamma^\ast) ] \} = 0$.

\begin{Schunk}
\begin{Sinput}
R> Grmst <- binregG(outrmtl, bmt, Avalues = 0:1)
R> summary(Grmst)
\end{Sinput}
\begin{Soutput}
G-estimator :
      Estimate Std.Err  2.5% 97.5%    P-value
risk0    15.04  0.7057 13.66 16.42 8.615e-101
risk1    17.45  1.8898 13.75 21.16  2.559e-20

Average Treatment effect: difference (G-estimator) :
   Estimate Std.Err   2.5% 97.5% P-value
pa    2.413   2.033 -1.572 6.398  0.2353

Average Treatment effect: ratio (G-estimator) :
log-ratio: 
    Estimate   Std.Err        2.5%     97.5%  P-value
pa 0.1487995 0.1190472 -0.08452883 0.3821277 0.211329
ratio: 
 Estimate      2.5%     97.5% 
1.1604402 0.9189452 1.4653993 
\end{Soutput}
\end{Schunk}

We here compute the G-estimates for the two-values given, thus are $15.0$ and $17.5$ for
\code{tcell} being $0$ or $1$, respectively. These are then compared with their difference and
ratio (on log-scale). 


The influence functions of the G-estimates are available for estimating the
covariance matrix, say,
\begin{Schunk}
\begin{Sinput}
R> crossprod(Grmst$risk.iid)
\end{Sinput}
\begin{Soutput}
            pa          pa
pa  0.49804182 -0.03254531
pa -0.03254531  3.57132760
\end{Soutput}
\end{Schunk}

To get the doubly-robust estimators we use the function  \code{mets::rmstATE}. This function
must have a factor as the variable that is used for the risk-estimates and 
subsequently for the ATE contrasts. This factor can have more than two-levels in which
case the propensity score model is a multinomial regression model (based on \code{mets::mlogit}).

\begin{Schunk}
\begin{Sinput}
R> dfactor(bmt) <- tcell.f~tcell
R> outrmst <- rmstATE(Event(time, cause != 0) ~ tcell.f + platelet + age,
+                     data = bmt, time = 30,
+                     treat.model = tcell.f ~ platelet + age,
+                     cens.model = ~ strata(platelet, tcell), model = "exp")
R> summary(outrmst)
\end{Sinput}
\begin{Soutput}
   n events
 408    231

 408 clusters
coeffients:
             Estimate   Std.Err      2.5%     97.5% P-value
(Intercept)  2.610518  0.057566  2.497692  2.723345  0.0000
tcell.f1     0.148799  0.119047 -0.084529  0.382128  0.2113
platelet     0.243180  0.084230  0.078092  0.408269  0.0039
age         -0.172771  0.037858 -0.246972 -0.098570  0.0000

exp(coeffients):
            Estimate     2.5%   97.5%
(Intercept) 13.60610 12.15440 15.2312
tcell.f1     1.16044  0.91895  1.4654
platelet     1.27530  1.08122  1.5042
age          0.84133  0.78116  0.9061

Average Treatment effects (G-formula) :
          Estimate  Std.Err     2.5%    97.5% P-value
treat0    15.04097  0.70572 13.65778 16.42416  0.0000
treat1    17.45415  1.88980 13.75022 21.15808  0.0000
treat:1-0  2.41318  2.03334 -1.57209  6.39844  0.2353

Average Treatment effects (double robust) :
          Estimate  Std.Err     2.5%    97.5% P-value
treat0    15.07860  0.70628 13.69433 16.46287  0.0000
treat1    16.96819  1.93472 13.17620 20.76017  0.0000
treat:1-0  1.88959  2.07098 -2.16946  5.94863  0.3616
\end{Soutput}
\end{Schunk}

The output first reports the outcome RMST regression model.
The  G-estimates are as before, and the doubly robust RMST estimates are
$15.1$ and $17.0$ for the factor \code{tcell},  respectively, and thus
quite similar to  those based on simple G-estimation. 

We can access the influence functions of the G-estimators and 
the augmented estimators directly from this output.
We here show the covariance matrix for the G-estimators and 
the double-robust estimators (DR)

\begin{Schunk}
\begin{Sinput}
R> crossprod(outrmst$riskG.iid)
\end{Sinput}
\begin{Soutput}
            riskGa.iid  riskGa.iid
riskGa.iid  0.49804182 -0.03254531
riskGa.iid -0.03254531  3.57132760
\end{Soutput}
\begin{Sinput}
R> crossprod(outrmst$riskDR.iid)
\end{Sinput}
\begin{Soutput}
            iidriska    iidriska
iidriska  0.49882483 -0.02349232
iidriska -0.02349232  3.74314326
\end{Soutput}
\end{Schunk}

Calling the \code{mets::rmstATE} with an outcome with multiple causes will lead
it to consider the RMTL for the specified cause (TRM, cause=1). We do not display
the output to save space. 

\begin{Schunk}
\begin{Sinput}
R> out1rmtl <- rmstATE(Event(time, cause) ~ tcell.f + platelet + age,
+                      data = bmt, time = 30, cause = 1,
+                      treat.model = tcell.f ~ platelet + age,
+                      cens.model = ~ strata(platelet, tcell),
+                      model = "exp")
\end{Sinput}
\end{Schunk}

We can access the different parts of the summary by applying the summary
function that has the components displayed

\begin{Schunk}
\begin{Sinput}
R> sout1rmtl <- summary(out1rmtl)
R> names(sout1rmtl)
\end{Sinput}
\begin{Soutput}
 [1] "coef"     "n"        "nevent"   "strata"   "ncluster" "var"     
 [7] "model"    "exp.coef" "ateG"     "ateDR"   
\end{Soutput}
\begin{Sinput}
R> sout1rmtl$ateG
\end{Sinput}
\begin{Soutput}
           Estimate   Std.Err      2.5%      97.5%      P-value
treat0    10.634753 0.7059923  9.251034 12.0184726 2.812505e-51
treat1     6.475311 1.5698900  3.398383  9.5522390 3.712327e-05
treat:1-0 -4.159442 1.7215168 -7.533553 -0.7853309 1.568562e-02
\end{Soutput}
\end{Schunk}

\section{Examples}\label{sec:examples}

In this Section we illustrate, considering the HF-ACTION study, 
how the influence functions at hand can  be utilized to compute derived measures. 
The HF-ACTION study \citep{OConnor2009} was a large, multi-center clinical
trial focused on patients with chronic heart failure.
It was designed to evaluate the role of structured exercise training in 
this population. The study specifically targeted patients with reduced 
ejection fraction.
Participants were recruited from multiple clinical centers across different
regions. They were randomly assigned to receive usual medical care or usual care
plus an exercise program. The intervention included both supervised and
home-based aerobic exercise. The trial was conducted under the name HF-ACTION.
Its main objective was to assess how exercise training influences clinical
outcomes. Researchers also aimed to examine adherence and safety of exercise in
heart failure patients. Overall, the study was designed to clarify the role of
exercise as part of standard heart failure management.
Patients were randomized to either
standard-of-care plus aerobic exercise training consisting of 36 supervised
sessions followed by home-based training, vs standard-of-care. Randomization was
stratified by heart failure etiology.

\subsection{Example I: RCT augmentation for RMST}

We consider an RCT set-up where the treatment $A$ is independent of other baseline covariates due to
randomization. We randomize with randomization probability $\Prob(A=1)=\pi$.
We add the additional assumptions that the distribution of $C| X,A$ only depends on $A$, and that in addition due to
randomization $A \indep X$. This implies the conditions of \cite{hattori2025sample},  namely that
$D \indep C | A$ and that $A \indep X$ that also considered the RCT augmentation approach for RMST. Their estimator was
based on solving the augmented equation
\begin{eqnarray*}
	U^{HT}(\hat \pi,\hat \gamma ,\mu(1)) &  = & \sum_i [ \frac{A_i}{\hat \pi}  \tilde O_i(\tau) - \frac{A_i - \hat \pi}{\hat \pi}  X_i^T \hat \gamma ] - \mu(1)  = 0
\end{eqnarray*}
and by finding $\gamma$ to minimize the variance. This is equivalent to regressing the influence functions of the simple RMST estimator on $(A-\pi) X_i$ or
regressing $A_i O_i(\tau) W(\tau)$ on $(A-\pi) X_i$. The estimator can also be written explicitely as
\begin{eqnarray*}
	\tilde \mu(1)  &  = & n^{-1} \sum_i \frac{A_i}{\hat \pi} \tilde O_i(\tau)  - \frac{A_i - \hat \pi}{\hat \pi} X_i^T \hat \gamma  \\
		       &  = & \int_0^\tau \hat S_1(s) ds  - n^{-1} \sum_i \frac{A_i - \hat \pi}{\hat \pi} X_i^T \hat \gamma
\end{eqnarray*}
where $\hat S_1(s)$ is the Kaplan-Meier estimator among the treated, again
using the equivalence between the IPCW and the Kaplan-Meier.

We note that we have the influence functions of the simple IPCW estimator directly from the \code{mets::rmstIPCW} function,
and in addition the G-estimator from the \code{mets::rmstATE} is equivalent to this estimator under our assumptions.

We note that in the RCT setting, $\E(D_\beta U)=0$ if the outcome model is found
by solving $\E_n [A(O(\tau) W(\tau) - m_1(X,\gamma))]=0$,
and that $\E(D_\gamma U)=0$ due to randomization
($X \indep Z$) and $\E(A)=\pi$. In this case the asymptotics are as
if $\pi$ and $\gamma$ are known.

Looking at the survival (\code{status=3}) in an RCT study, the HF-ACTION study \citep{OConnor2009}, we start by computing the RMST for those treated and those not.
We also make some covariates that can be used to improve precision if they would be predictive for the outcome (in this case not due to construction).

\begin{Schunk}
\begin{Sinput}
R> data(hfactioncpx12)
R> dtable(hfactioncpx12, ~status)
\end{Sinput}
\begin{Soutput}

status
   0    1    2 
 617 1391  124 
\end{Soutput}
\begin{Sinput}
R> hf <- hfactioncpx12
R> cid <- countID(hf)
R> hflast <- subset(hf, cid$reverseCountid==1)
R> set.seed(1)
R> hflast$time <- hflast$time + runif(nrow(hflast))*.001 # break ties
R> dtable(hflast, ~status)
\end{Sinput}
\begin{Soutput}

status
  0   2 
617 124 
\end{Soutput}
\begin{Sinput}
R> hflast$age <- rnorm(nrow(hflast))
R> hflast$sex <- rbinom(nrow(hflast), 1, 0.5)
R> 
R> outrct <- rmstATE(
+    Event(time, status != 0) ~ treatment +age:treatment +sex:treatment,
+    data = hflast, time = 3,
+    treat.model = treatment ~ 1,
+    model = "lin", cens.model =~ strata(treatment)
+  )
R> summary(outrct)
\end{Sinput}
\begin{Soutput}
   n events
 741    114

 741 clusters
coeffients:
                 Estimate    Std.Err       2.5%      97.5% P-value
(Intercept)     2.6715341  0.0523211  2.5689867  2.7740815  0.0000
treatment1      0.1223514  0.0681988 -0.0113159  0.2560187  0.0728
treatment0:age -0.0614252  0.0415520 -0.1428657  0.0200152  0.1393
treatment1:age  0.0152211  0.0326542 -0.0487800  0.0792222  0.6411
treatment0:sex -0.0125054  0.0785909 -0.1665407  0.1415299  0.8736
treatment1:sex  0.0087238  0.0629041 -0.1145660  0.1320135  0.8897


Average Treatment effects (G-formula) :
          Estimate  Std.Err     2.5%    97.5% P-value
treat0    2.668564 0.039053 2.592022 2.745107  0.0000
treat1    2.797293 0.030664 2.737193 2.857394  0.0000
treat:1-0 0.128729 0.049684 0.031351 0.226108  0.0096

Average Treatment effects (double robust) :
          Estimate  Std.Err     2.5%    97.5% P-value
treat0    2.668564 0.039053 2.592022 2.745107  0.0000
treat1    2.797293 0.030664 2.737193 2.857394  0.0000
treat:1-0 0.128729 0.049684 0.031351 0.226108  0.0096
\end{Soutput}
\end{Schunk}

The doubly robust estimates suggest that the RMST is $2.8 (2.8,2.9)$ for those
exercising and $2.6 (2.5,2.8)$ for those on standard care, on the RMST scale the
ATE is $0.1 (0,0.2)$. Patients receiving standard care lost, on average, 0.1
year of restricted mean survival time. The G-estimates and the doubly robust
estimates are equivalent in this case and are potentially more efficient than
the simple RCT estimates by taking advantage of baseline covariates.


%
%
%


\subsection{Example II: While alive influence function fun}

In the recurrent events setting consider the while-alive estimand, the expected number of events up to time $\tau$ 
divided with the RMST 
\begin{eqnarray*}
	\frac{\E[N^r( D \wedge \tau)]}{\E( D \wedge \tau)}
\end{eqnarray*}
where $N^r(t)$ counts the recurrent events up to time $t$, and $D$ is the terminal event (the time of death, for example),
see for example \cite{mao2023,schmidli2021estimands}. Both recurrent events and $D$ can be observed with independent right-censoring, as described in
\cite{mao2023}.

We estimate $\E[N^r(D \wedge \tau) | A=j]$ for $j=0,1$, where $A$ is a binary treatment indicator, using IPCW regression for the
outcome $N^r(D \wedge \tau)$ and the RMST regression for $D \wedge \tau$ as described earlier. 
We will not go into details of the recurrent events regression but the key is just that we can fit such a regression model 
and that the influence functions are available. The ratio of the two IPCW estimators is equivalent to the estimator 
suggested in \citep{mao2023} that is also implemented in 
the package \pkg{WA} \citep{wa-r}. 
In \pkg{mets} this estimator is also available via the \code{mets::WA\_recurrent} function.

\begin{Schunk}
\begin{Sinput}
R> dtable(hfactioncpx12, ~status)
\end{Sinput}
\begin{Soutput}

status
   0    1    2 
 617 1391  124 
\end{Soutput}
\begin{Sinput}
R> hf <- hfactioncpx12
R> cid <- countID(hf)
R> 
R> hflast <- subset(hf, cid$reverseCountid == 1)
R> dtable(hflast, ~status)
\end{Sinput}
\begin{Soutput}

status
  0   2 
617 124 
\end{Soutput}
\begin{Sinput}
R> outN <- recregIPCW(
+    Event(entry, time, status) ~ -1 + treatment + cluster(id),
+    data = hf,time = 3, cause = 1,
+    cens.model = ~strata(treatment), model="lin"
+  )
R> summary(outN)
\end{Sinput}
\begin{Soutput}
   n events
 741   1281

 741 clusters
coeffients:
           Estimate Std.Err    2.5%   97.5% P-value
treatment0  2.11850 0.11385 1.89535 2.34164       0
treatment1  1.92406 0.12165 1.68564 2.16248       0
\end{Soutput}
\begin{Sinput}
R> head(IC(outN))
\end{Sinput}
\begin{Soutput}
         [,1]     [,2]
1  0.33659716 0.000000
2  0.72299325 0.000000
3  0.00000000 6.892408
4 -2.19842373 0.000000
5 -0.08304862 0.000000
6 -5.23842328 0.000000
\end{Soutput}
\begin{Sinput}
R> outD <- rmstIPCW(
+    Event(time, status == 2) ~ -1 + treatment + cluster(id),
+    data = hflast, time = 3,
+    cens.model = ~ strata(treatment), model="lin"
+  )
R> summary(outD)
\end{Sinput}
\begin{Soutput}
   n events
 741    114

 741 clusters
coeffients:
           Estimate  Std.Err     2.5%    97.5% P-value
treatment0 2.669249 0.038983 2.592845 2.745654       0
treatment1 2.797565 0.030593 2.737604 2.857527       0
\end{Soutput}
\begin{Sinput}
R> head(IC(outD))
\end{Sinput}
\begin{Soutput}
         [,1]      [,2]
1  0.31492608 0.0000000
2  0.76941871 0.0000000
3  0.00000000 0.1526227
4 -5.19264265 0.0000000
5  0.01352951 0.0000000
6  0.76941871 0.0000000
\end{Soutput}
\end{Schunk}

The first set of estimates from the recurrent events regression, shows
that the treatment group has $1.9 (1.7,2.2)$ events on average up to 3 years, and
an RMST of $2.8 (2.7,2.9)$. The ratio of these two estimates is
thus the average number of events relative to the time surviving (out to 3 years).
Getting the standard errors is more complex and we here estimate these
taking advantage of the influence functions of the two sets of estimates. 

Using the \pkg{lava}-package we can combine the influence functions,
a \code{cbind}-call would also do in this case since the influence functions are
ordered after the ``id'' in the \code{hf} data.
Using the influence functions we
can therefore estimate the while-alive ratio and its standard errors, and
make a test for equivalence between the treated and un-treated.

\begin{Schunk}
\begin{Sinput}
R> library(lava)
R> eD <- estimate(outD)
R> eN <- estimate(outN)
R> eND <- merge(eD, eN)
R> 
R> ratio <- estimate(eND, function(p) c(p[3] / p[1], p[4] / p[2]))
R> ratio
\end{Sinput}
\begin{Soutput}
             Estimate Std.Err   2.5%  97.5%   P-value
treatment0.1   0.7937 0.04464 0.7062 0.8812 1.051e-70
treatment1.1   0.6878 0.04474 0.6001 0.7755 2.535e-53
\end{Soutput}
\begin{Sinput}
R> estimate(ratio, rbind(c(1,-1)))
\end{Sinput}
\begin{Soutput}
                          Estimate Std.Err     2.5%  97.5% P-value
[treatment0.1] - [tre....   0.1059 0.06321 -0.01798 0.2298 0.09383
------------------------------------------------------------
Null Hypothesis: 
  [treatment0.1] - [treatment1.1] = 0 
 
chisq = 2.8075, df = 1, p-value = 0.09383
\end{Soutput}
\end{Schunk}

The ratio is thus $0.8 (0.7,0.9)$ for those on standard care,  and the two
ratios are not significantly different ($p=0.1$).

We have illustrated how the influence functions are useful for combining different estimates, but  the while-alive ratio
can be computed directly in the  \code{mets::WA_recurrent} function that computes the ratio as well as the mean number of events per time-unit
\cite{ragni2024patient}

\begin{Schunk}
\begin{Sinput}
R> e <- WA_recurrent(Event(entry, time, status) ~ treatment + cluster(id),
+                    data = hf, time = 3, death.code = 2) 
R> summary(e) 
\end{Sinput}
\begin{Soutput}
While-Alive summaries:  

RMST,  E(min(D,t)) 
           Estimate Std.Err  2.5% 97.5% P-value
treatment0    2.669 0.03898 2.593 2.746       0
treatment1    2.798 0.03059 2.738 2.858       0
 
                          Estimate Std.Err    2.5%    97.5%  P-value
[treatment0] - [treat....  -0.1283 0.04955 -0.2254 -0.03119 0.009614
------------------------------------------------------------
Null Hypothesis: 
  [treatment0] - [treatment1] = 0 
 
chisq = 6.7051, df = 1, p-value = 0.009614
mean events, E(N(min(D,t))): 
           Estimate Std.Err  2.5% 97.5%   P-value
treatment0    2.118  0.1139 1.895 2.342 2.792e-77
treatment1    1.924  0.1216 1.686 2.162 2.381e-56
 
                          Estimate Std.Err    2.5% 97.5% P-value
[treatment0] - [treat....   0.1944  0.1666 -0.1321 0.521  0.2432
------------------------------------------------------------
Null Hypothesis: 
  [treatment0] - [treatment1] = 0 
 
chisq = 1.3618, df = 1, p-value = 0.2432
_______________________________________________________ 
Ratio of means E(N(min(D,t)))/E(min(D,t)) 
   Estimate Std.Err   2.5%  97.5%   P-value
p1   0.7937 0.04464 0.7062 0.8812 1.051e-70
p2   0.6878 0.04474 0.6001 0.7755 2.535e-53
 
            Estimate Std.Err     2.5%  97.5% P-value
[p1] - [p2]   0.1059 0.06321 -0.01798 0.2298 0.09383
------------------------------------------------------------
Null Hypothesis: 
  [p1] - [p2] = 0 
 
chisq = 2.8075, df = 1, p-value = 0.09383
_______________________________________________________ 
Mean of Events per time-unit E(N(min(D,t))/min(D,t)) 
       Estimate Std.Err   2.5%  97.5%   P-value
treat0    1.083 0.12300 0.8419 1.3241 1.312e-18
treat1    0.737 0.06395 0.6116 0.8623 9.911e-31
 
                    Estimate Std.Err    2.5%  97.5% P-value
[treat0] - [treat1]    0.346  0.1386 0.07432 0.6177 0.01256
------------------------------------------------------------
Null Hypothesis: 
  [treat0] - [treat1] = 0 
 
chisq = 6.2304, df = 1, p-value = 0.01256
\end{Soutput}
\end{Schunk}

In the HF-ACTION study, the exercise group
versus the standard care group
showed a ratio of means at $0.69$ vs $0.79$ ($p=0.094$),
and means of events per time unit 
at  $0.74$ vs $1.08$ ($p=0.013$).
The means of events per time unit has the advantage that
it takes into account 
the correlation between event counts and
individual time-at-risk thus
focusing on the individual 
burden relative to exposure. 

\section[cif-regression]{Binomial regression for competing risks\label{sec:cif-reg}}

The key focus has been on the RMST and RMTL outcomes, but as already indicated another
important outcome is the $O_j(\tau) = I(D \leq \tau, \eta=j)$, and now
the $\E(O_j(\tau) | X )= \Prob(D \leq \tau, \eta=j | X) = F_j(\tau,X)$, the cumulative
incidence given $X$, \citep{Scheike2008a,Blanche2022}.

This outcome is also considered in the \pkg{mets} package, and we have 
the same functionality as described for the outcomes we focused on this presentation.

The IPCW estimators are equivalent to the Aalen-Johansen estimators for a stratified
covariate, and we can do regression using the
\code{mets::binreg} function, and compute the ATE using \code{mets::binregATE} function.

\section{Discussion}\label{sec:discussion}

%
%

This paper presents a comprehensive software implementation within the mets
R-package for analyzing Restricted Mean Survival Time (RMST) and Restricted Mean
Time Lost (RMTL), including their decomposition in competing risks settings.
While RMST has gained traction as a clinically interpretable alternative to
hazard ratios, our contribution distinguishes itself through a unique technical
capability: the simultaneous computation of non-parametric estimates and their
standard errors across all time horizons. However, the most significant
advancement offered by this work is the systematic provision of influence
functions for all implemented models, ranging from non-parametric estimators to
complex regression frameworks based on Inverse Probability of Censoring
Weighting (IPCW).

The availability of these influence functions transforms the \pkg{mets} package from a
mere estimation tool into a flexible building block for advanced statistical
inference. By returning the influence functions for every model, we enable users
to construct custom statistics that were previously difficult to derive
analytically. For instance, we demonstrate how these functions facilitate the
calculation of the "while-alive" estimand in recurrent events settings, a ratio
of event rates to survival time that requires combining variances from distinct
estimators. Furthermore, the influence functions are critical for implementing
doubly robust estimators for Average Treatment Effects (ATE) and G-computation,
ensuring valid standard errors even when models are partially misspecified.

Unlike other packages that rely on computationally intensive bootstrapping for
variance estimation, our approach leverages these influence functions to achieve
linear scaling with the number of observations, making it highly efficient for
large-scale clinical trials. Crucially, this framework naturally extends to data
with complex dependency structures. As highlighted in our final section, when
observations are clustered, such as in multi-center trials or family studies,
the
availability of individual-level influence functions allows for the direct
computation of Generalized Estimating Equation (GEE) type standard errors. By
simply aggregating the influence functions within clusters, we can obtain robust
variance estimates that account for intra-cluster correlation without resorting
to resampling methods. This capability ensures that the rigorous inference
provided by our RMST and RMTL models remains valid and efficient even in the
presence of hierarchical data structures, solidifying the \pkg{mets} package as a
versatile tool for modern biostatistical analysis.

\section*{Acknowledgement}
We appreciate comments from Takahiro Hasegawa.

\FloatBarrier%
\newpage
\bibliography{ref}

\end{document}